\documentclass[sts, noinfoline]{imsart_Arxiv}

\startlocaldefs
\endlocaldefs

\usepackage{color}
\usepackage{xcolor}
\usepackage{graphicx} 
\graphicspath{ {./Figures/} }
\usepackage{subcaption}
\usepackage[authoryear, round]{natbib} 
\usepackage{hyperref}
\usepackage{comment}
\usepackage{rotating}
\usepackage{afterpage}

\begin{document}

\begin{frontmatter}

\title{Variability in the interpretation of Dutch probability phrases - a risk for miscommunication}
\runtitle{Variability in interpretation of Dutch probability phrases} 

\begin{aug}
	\author{\fnms{Sanne J.W.} \snm{Willems}
		\ead[label=e1]{s.j.w.willems@math.leidenuniv.nl}},
	\author{\fnms{Casper J.} \snm{Albers}\ead[label=e2]{c.j.albers@rug.nl}}
	\and
	\author{\fnms{Ionica} \snm{Smeets}
		\ead[label=e3]{i.smeets@biology.leidenuniv.nl}
}
	
	\runauthor{S.J.W. Willems et al.}
	
	\affiliation{Leiden University and University of Groningen, The Netherlands}
	
	\address{Sanne J.W. Willems is PhD candidate, Mathematical Institute, Leiden University, 
		2333 CA Leiden 
		 \printead{e1}.}
	\address{Casper J. Albers is Professor, Department Psychometrics \& Statistics, University of Groningen, 
		9712 TS Groningen 
		 \printead{e2}.}
	
	\address{Ionica Smeets is Professor, Department Science Communication and Society, Leiden University, 
		2333 BE Leiden 
		 \printead{e3}.}
\end{aug}

\begin{abstract}
Verbal probability phrases are often used to express estimated risk. In this study, focus was on the numerical interpretation of 29 Dutch probability and frequency phrases, including several complementary phrases to test (a)symmetry in their interpretation. Many of these phrases had not been studied before. The phrases were presented in the context of ordinary situations. The survey was distributed among both statisticians and non-statisticians with Dutch as their native language.
	
The responses from 881 participants showed a large variability in the interpretation of Dutch phrases, and the neutral contexts seemed to have no structural influence. Furthermore, the results demonstrated an asymmetry in the interpretation of Dutch complementary phrases. 
The large variability of interpretations was found among both statisticians and non-statisticians, and among males and females, however, no structural differences were found between the groups.

Concluding, there is a large variability in the interpretation of verbal probability phrases, even within sub-populations. Therefore, verbal probability expressions may be a risk for miscommunication.

\end{abstract}


\end{frontmatter}

\section{Introduction}
Every day people make decisions based on estimated probabilities and risks. These decisions range from choices with little consequences (should I bring my umbrella to avoid getting wet in the rain?) to important life decisions (which treatment will most likely cure my cancer without causing too many undesirable side-effects?). Many of our decisions rely on risks expressed by others (weather forecasters or oncologists). Due to this dependence of the decision maker on the information provider, it is important that the message is conveyed as intended in order to minimize miscommunication.

Research has shown that information providers, the senders of a message, prefer to express probabilities verbally, i.e.\ by using phrases as {\em unlikely\/}, {\em usually} and {\em maybe\/}, while decision makers favor numeric expressions like percentages. \citet{Art:PreferenceParadox_ErevCohen} refer to this as the {\em communication mode preference paradox\/}.

\citet{TechRep:LiteratureReview_Druzdzel} reasoned that senders prefer verbal expressions because these convey some amount of uncertainty. Including this uncertainty in the expression is favored by senders, because probability estimates are usually based on empirical data and therefore not sufficiently precise to be translated into exact numerical statements. Hence, if a numerical value is given, its suggested precision may be misleading. On the other hand, decisions makers prefer this precision of numerical expressions, since numeric values are easier to compare and to draw conclusions from. 

Due to the conflicting preferences for the mode of communication, a translation step from the verbal phrases to numeric values (and vice versa) is required in the communication process. To avoid miscommunication in this translation step, some researchers intended to develop a translation table. 

As a first step towards making such a codification, studies were done to examine the interpretation of probability phrases. The designs of these studies were very comparable; respondents were asked to give their interpretation of a probability expression as a single value or range on a scale of 0-1 or 0-100, or were asked to rank them. The phrases were either presented out-of-context or in sentences describing a particular situation. Many of these studies were summarized in the literature reviews by \citet{TechRep:LiteratureReview_Druzdzel} and \citet{Art:LiteratureReview_VisschersEtAl} and the meta-analysis by \citet{Art:Meta-Analysis_Theil}.

The overall conclusion from these studies was that, although individuals seem to be internally consistent in their ranking of probability phrases \citep{Art:ConsistencyInInterpretation_BudescuWallsten} and their perception of them over time \citep{Art:ExpressionsOfProb_BryantNorman}, the interpretation of these phrases varies greatly among individuals.
This interpretation variability is especially large for phrases expressing a probability in the range from 20\% to 80\%. For words that express extreme probabilities, e.g.\ {\em always\/}, {\em certain\/}, {\em never\/} and {\em impossible\/}, consensus was highest. Additionally, the numerical interpretations of some probability phrases were very similar and, therefore, some expressions can be considered to be synonyms. For example, \citet{Art:QuantMeaningProbExpr_ReaganYoutz1989} concluded that {\em likely\/} is synonymous with {\em probable\/} and {\em low chance\/} with {\em unlikely\/} and {\em improbable\/}. 

Based on the discovery of this synonymous pair, \citet{Art:QuantMeaningProbExpr_ReaganYoutz1989} also expected that {\em high chance\/} would be synonymous with {\em likely\/} and {\em probable}. However, their data indicated that actually {\em very likely\/} and {\em very probable\/} are its synonyms. This unbalanced result shows that there is some asymmetry in the interpretation of probability phrases. 

The asymmetry in the interpretation of mirrored probability phrases is a phenomenon studied and confirmed by many researchers. For example, \citet{Art:EmpiricalScaling_LichtensteinNewman} focussed on the influence of adverbs (e.g.\ {\em very\/}, {\em quite\/} and {\em fairly\/}) attached to mirrored phrases as {\em likely\/} and {\em unlikely\/}. They found that, for instance, the mean of the numeric probability given to {\em quite likely\/} was 79\%, while the mean for {\em quite unlikely\/} was 11\%. These means sum up to 90\% instead of 100\%, indicating an asymmetry in the interpretation of the two complementary expressions. 

There are some mirrored terms that show a very strong asymmetry in their interpretation. For example, \citet{Art:QuantifyingProbExpr_MostellerYoutz} studied the terms {\em possible\/} and {\em impossible\/}. According to their data, the interpretation of {\em impossible\/} is stable (around 3\% for all participants of the study), while {\em possible\/} has distinct meanings for different groups. Namely, some respondents used the literal interpretation of {\em possible\/} and indicated that it could indicate any percentage between 0\% and 100\%, and others associated it with rare events that only scarcely occur (as in {\em barely possible\/}). Hence, the different interpretations of {\em possible\/} causes the strong asymmetry with its mirrored expression {\em impossible\/}.

All these results show that the interpretations of verbal probability expressions vary too much to translate them into numerical values which would be supported by everyone. Therefore, many researchers who initially intended to make a translation table, concluded that such a codification is probably impossible (e.g.\ \citet{Art:QuantifyingProbExpr_MostellerYoutz, Art:LostForWords_TimmermansMileman, Art:EmpiricalScaling_LichtensteinNewman, Art:ContextualEffects_WeberHilton}), or realized that their currently used table was actually not conveying the intended probabilities \citep{Art:FormuleringInfoBijwerkingen_PanderMaat}.

\section{Aims of this study}

This study was focussed on the variability of the numerical interpretation of Dutch probability phrases. There have been several previous studies in Dutch, but these mainly concentrated on expressed frequencies and just a few probability phrases. To fill this gap on probability expressions, our survey included many phrases, expressing both probabilities and frequencies. We compared the interpretations of these phrases found in our study with the interpretations of those in other studies in Dutch and with the interpretations of their English translations found in English studies. Additionally, we studied synonymous phrases and asymmetric expressions, since these have not yet been analysed in previous Dutch studies. 

In contrast to previous studies, in which phrases were placed out-of-context or in a specific context of interest, we presented phrases in a neutral contexts based on ordinary events. In this way, we tried to investigate whether the interpretation variability is also high when all participants use a similar context that is less susceptible to prior beliefs. 

Furthermore, previous researchers suggested that the interpretations of phrases may differ between groups. In our study, we investigated this hypothesis further by comparing the interpretations of statisticians with those of non-statisticians, and by comparing the results from male and female respondents.

\subsubsection*{Previous Dutch  studies}
A lot of research on the numerical interpretation of probability phrases was conducted in English. For example, all the studies listed in the introduction focussed on English phrases. Additionally, there have been replication studies in Dutch (and many other languages). \citet{Art:DoorgaansDikwijls_EekhofMolPielage} studied the interpretation of 30 Dutch phrases. However, these phrases expressed frequencies instead of probabilities. In a study by \citet{Art:DutchPhrases_Timmermans} some probability phrases were included, usually in combination with an adverb like {\em quite\/} or {\em rather\/}. Unfortunately, the article is written in English and does not provide the Dutch expressions used in the study, hence it is unclear exactly which Dutch expressions and adverbs were investigated. \citet{Art:TalkingProb_RenooijWitteman} did several experiments to develop a probability scale containing both words and numbers. Their focus was on ranking seven probability phrases and developing their corresponding numerical scale. In a study by \citet{Art:FormuleringInfoBijwerkingen_PanderMaat}, focus was on the interpretation of uncertainty in information leaflets that come with medicine. Although their main interest was not in the numerical values associated with verbal probability phrases, they did investigate this for three phrases. 

Given that the first study included many phrases but only frequencies, and the other three studies included only a few probability phrases, usually in combination with adverbs, many Dutch probability expressions still needed to be studied. Therefore, we focussed on probability phrases with and without adverbs and, to compare with the results found by \citet{Art:DoorgaansDikwijls_EekhofMolPielage}, included several frequency expressions as well.

\subsubsection*{Translating probability phrases}
In addition to replication studies in other languages, several studies have been done to compare the interpretation variability of English probability phrases with the interpretations of their translations to other languages. Three studies, comparing English with French \citep{Art:FrenchVSEnglish_DavidsonChrisman}, German \citep{Art:GermanVSEnglish_DoupnikRichter}, and Chinese \citep{Art:ChineseVSEnglish_HarrisEtAll}, showed that on average the numerical interpretations of the English phrases differ from the interpretation of their counterparts in the three other languages. Additionally, in French and Chinese, the standard deviations of the numerical values related to the probability phrases were much larger than those of the original English wording. 

These results show that the meaning of probability expressions can get lost in translation from one language to another. This problem is especially relevant for international or intergovernmental documents. Namely, when these documents are translated from one language to another, it is important that the rewording of probability phrases communicate the same quantifications of uncertainty and probability as intended in its original language. For the three studies mentioned before, Canadian governmental documents, the International Accounting Standards, and documents from the Intergovernmental Panel on Climate Change were used to compare English with French, German, and Chinese respectively. Based on the results, we may conclude that probably not all translations of these documents convey the same probability messages. To find out whether this problem may also occur when translating English to Dutch, or visa versa, we also compared the interpretations of some Dutch phrases with their English translations included in other studies.

\subsubsection*{Synonyms and asymmetry}
As mentioned in the introduction, several studies on the interpretation of English probability phrases \citep{Art:EmpiricalScaling_LichtensteinNewman, Art:QuantMeaningProbExpr_ReaganYoutz1989, Art:BlindChance_Stheeman} have investigated whether some expressions are synonyms and whether there is symmetry in the interpretations. These concepts were, however, not researched in the previous Dutch studies. Therefore, we will investigate them in this study and compare them to synonymous and complementary phrases in English.

\subsubsection*{Influence of context}

An important finding from previous studies was that the interpretation of a probability phrase is influenced enormously by its context. For instance, compare your numerical interpretation of the word {\em likely\/} in the next two statements:
\begin{itemize}
	\item {\em It is likely that it will rain in Manchester, England, next June;}
	\item {\em It is likely that it will rain in Barcelona, Spain, next June.}
\end{itemize}
Probably, your numerical interpretation of {\em likely\/} in the first statement is higher than in the second. \citet{Art:BaseRateEffect_WallstenEtAl} used this example and, based on their research, predicted a difference in their numerical interpretation. Namely, in their study, they showed that an individual's expected base-rate of a context scenario influences this person's interpretation of the probability phrase. In this example the base-rate for the first scenario is higher, since in general it is more likely to rain in England than in Spain in June, and this influences the interpretation of the word {\em likely\/}.

This hypothesis on the base-rate effect was confirmed by \citet{Art:ContextualEffects_WeberHilton}, who, additionally, provided evidence that other variables may be affecting the interpretation as well. According to their findings, the perceived severity or consequentiality of an event and its emotional valence will also influence the judged probability.

Since it was shown that context may influence the interpretation of probability phrases, many researchers decided to investigate them out-of-context. However, it was argued by \citet{TechRep:LiteratureReview_Druzdzel} that, if no specific context is provided, participants may invent their own context. Due to these self-created contexts, participants' responses will portray the interpretation of the probability phrases in many completely different contexts instead of out-of-context. These different scenarios may cause extra variability in the data which makes it more difficult to draw conclusions from the results. 

This is the reason why we decided to not present the phrases out-of-context. However, to reduce the influence of prior beliefs, we tried to choose contexts that were neutral, i.e.\ ordinary situations that everyone can relate to, but which do not induce strong prior expectations.

\subsubsection*{Differences between sub-populations}
In most studies, data on the interpretation of probability phrases was gathered within specific sub-populations. Participants were, for instance, physicians \citep{Art:ExpressionsOfProb_BryantNorman}, science writers \citep{Art:QuantifyingProbExpr_MostellerYoutz}, radiologists \citep{Art:BlindChance_Stheeman}, biological scientists \citep{Art:CommunicatingPlantRisk_MacLeod}, or patients \citep{Art:FormuleringInfoBijwerkingen_PanderMaat}. Although all these studies showed variability in the perception of probability phrases within these sub-populations, one might wonder whether there are any differences between these groups as well. For example, \citet{Art:Meta-Analysis_Theil} argued that there may be a difference between professionals, who regularly make and communicate probability estimations, and persons who are inexperienced in this respect. However, his meta-analysis did not provide evidence for this hypothesis.  

On the other hand, in studies on the use of jargon, it has been shown that there is a significant difference in the interpretation of medical terms between doctors and patients \citep{Art:JargonMedical_Boyle} and of hydrological vocabulary between experts and laymen \citep{Art:JargonGeosciences_VenhuizenAlbersSmeets}. Experts may be unaware of this difference \citep{Art:JargonMedical_Castro} and, hence, their use of jargon may cause a miscommunication of information. 

Given these results on the different interpretations of jargon, there is reason to believe that there may be differences between the numerical interpretations of probability expressions of experts and laymen as well, as \citet{Art:Meta-Analysis_Theil} suggested. If this hypothesis is correct, experts may be misunderstood if they express probabilities verbally. Therefore, we decided to investigate this hypothesis further and distributed our survey among both statisticians and non-statisticians. This allowed us to compare the interpretations of these two groups. 

Additionally to comparing experts with laymen, one could also wonder whether there are differences between genders. To our knowledge, so far no study was done in which this comparison was made. Therefore, we investigated gender effects as well.

\section{Survey development and data collection}

The survey design for this study was based on the survey set-up of previous studies. In short, probability phrases were presented to participants, and they could give their interpretation as a point estimate on a 0-100\% scale. 

\subsubsection*{Choice of phrases}
There are many Dutch probability and frequency phrases that can be studied. To make a selection for our study, we first listed the phrases used in the English studies and translated them to Dutch using {\em Google Translate\/} \citep{GoogleTranslate} and the leading Dutch dictionary {\em Van Dale\/} \citep{VanDale}. Then we added the expressions from previous Dutch studies \citep{Art:DoorgaansDikwijls_EekhofMolPielage, Art:TalkingProb_RenooijWitteman, Art:FormuleringInfoBijwerkingen_PanderMaat}. This resulted in a list of 131 phrases.

This list was too long to use in one survey, so a selection had to be made. Since it makes sense to investigate the most frequently used phrases, we selected the phrases that were used at least 100 times in headlines of the popular Dutch news website {\em nu.nl\/}. To prevent too much overlap with the research by \citet{Art:DoorgaansDikwijls_EekhofMolPielage}, only the ten most commonly used frequency phrases were selected. Furthermore, some combinations of adverbs with a probability phrase were removed from the list to prevent too much overlap with the study by \citet{Art:DutchPhrases_Timmermans}, and to prevent repeats of very similar phrases. Additionally, the word {\em undecided\/} was removed, since it was mostly used in the headlines of sport results, in which it has a different meaning. 

This method of phrase selection resulted in a list of 29 frequency and probability expressions. These phrases, and their English translations, are given in Table~\ref{PPhr_tab:TranslationProbabilityPhrases} in Appendix~\ref{PPhr_app:TranslationProbabilityPhrases}. In this article, we will use the English translations. Please keep in mind that all given numerical interpretations for these phrases are actually for their Dutch counterparts.

\subsubsection*{Context}
As described before, the interpretation of a probability expression may be influenced by a person's prior expectations of the phrase's context. To avoid these base-rate effects, our aim was to formulate sentences that were neutral in the sense that everyone can imagine the situation, but has no prior expectations about it. Some examples of the sentences, formulated with the probability phrase {\em likely\/}, are
\newpage
\begin{itemize}
	\item {\em It is \textbf{likely} that this plan succeeds.}
	\item {\em It is \textbf{likely} that this hotel is fully booked.}
	\item {\em It is \textbf{likely} that the team wins a match.}
\end{itemize}
The probability phrases were printed in bold to put more focus on them. We tried to minimize the base-rate effect by not specifying a specific plan, hotel, or team. We developed twelve sentences like these. The complete list of these contexts is given in Table~\ref{PPhr_tab:ContextSentencesUsed} in Appendix~\ref{PPhr_app:ContextPhrases}. 

\subsubsection*{Numeric interpretations}
For each probability expressions in the survey, participants gave the point estimate of their numerical interpretation in percentages (0\% - 100\%) by using a slider. For example, the questions related to the three statements above were formulated as follows:
\begin{itemize}
	\item {\em What is the probability (expressed in percentages) that this plan succeeds?}
	\item {\em What is the probability (expressed in percentages) that this hotel is fully booked?}
	\item {\em What is the probability (expressed in percentages) that the team wins a match?}
\end{itemize}
All probability phrases were presented individually and in a random order, and participants were required to answer each question before continuing to the next. These forced-choices represent the real-life situation in which you also have to interpret each verbal probability. Additionally, in this way, missing data was prevented.

\subsubsection*{Randomization}
To prevent a systematic influence of the context on the interpretation of the probability phrase, 12 different versions of the survey were created. In every version, the probability phrase was formulated in a different context sentence and contexts were repeated two or three times in each survey version (since 29 is not divisible by 12). All survey versions were evenly and randomly distributed among the participants by the survey software \citep{Qualtrics}.

\subsubsection*{Personal characteristics}
After giving their interpretation of the 29 phrases, participants were asked for some personal information. This included their native language, whether they are a statistician, their highest completed education level, age, and gender. All these were categorized. For native language, participants could choose either {\em Dutch\/}, {\em Flemish\/}, or {\em Other\/}. Being a statistician was questioned as {\em``Are you a statistician or do you perform statistical analyses on a weekly or monthly basis?"\/}. Education was categorized in six common categories of degrees in the Netherlands. Age was categorized in intervals of 20 years and participants were allowed to refrain from providing their age. Gender was categorized as {\em male\/}, {\em female\/}, and {\em other/prefer not to say\/}. All these characteristics were asked as multiple choice questions and participants could select one of the given categories. 

\subsubsection*{Pilot}
A pilot study showed that the length of the survey was reasonable (approximately ten minutes) and that the explanation was clear. We noticed that some participants had the tendency to base their interpretation of a phrase on their interpretations of previous phrases. This confirmed that randomisation of the phrases is necessary. Additionally, it supported our decision to present one phrase at the time and to not allow participants to change their answers. If we would have permitted this, participants may have ranked their answers instead of giving the interpretations individually, which may have influenced the results. Based on the pilot study, we decided to make the original question {\em``Are you a statistician or do you perform statistical analyses on a regular basis?"\/} more specific by changing {\em ``on a regular basis"\/} into {\em``on a weekly or monthly basis?"\/}. 

\subsubsection*{Survey distribution}
We obtained permission to distribute this survey from the ethical committee of the Faculty of Behavioural and Social Sciences of the University of Groningen (17451-O). Since we wanted to compare Dutch-speaking statisticians with non-statisticians, the survey was distributed among both groups. Statisticians were invited via the mailing list of the Netherlands Society for Statistics and Operations Research (VVSOR) and the Interuniversity Graduate School of Psychometrics and Sociometrics (IOPS). To reach non-statisticians, the survey invitation was distributed via Twitter \citep{Twitter}.

\section{Participants' characteristics}

The survey was open for participation for almost four months, namely between July 18th 2018 and November 8th 2018. During this time, $1004$ persons started the survey, of which $115$ did not finish it. These incomplete observation were removed from the data. Another $8$ participants were removed from the data, because their native language was not Dutch or Flemish. As a result, data from 881 participants was used for analysis.

Flemish is formally a Dutch dialect and it was the native language of only seven participants. Since this group was too small to analyse separately, we merged categories {\em Dutch\/} and {\em Flemish\/}. This decision to merge was supported by the research conducted by \citet{Art:GermanVSEnglish_DoupnikRichter} who studied German-speaking participants from Germany, Austria and Switzerland and found no nationality-effect. 

\begin{figure*}[h]
	\centering
	\includegraphics[width=0.95\textwidth]{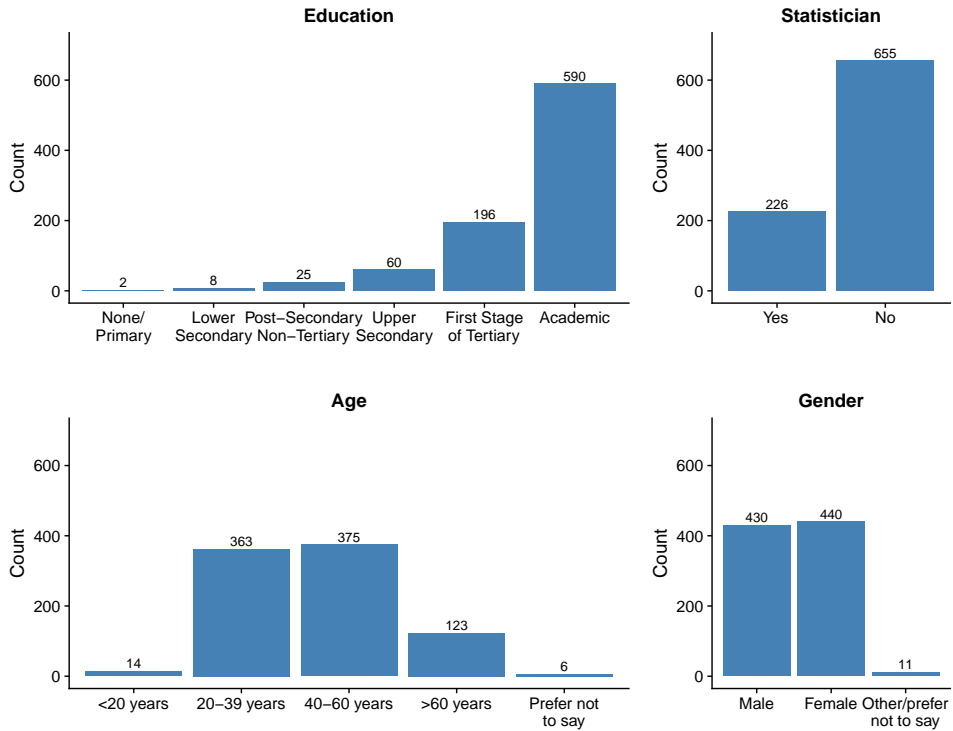}
	\caption{Bar graphs of the number of participants in each of the category levels of variables {\/\em Education}, {\/\em Statistician}, {\/\em Age}, and {\/\em Gender}.}
	\label{PPhr_fig:HistorgramsDistributionCatVar}
\end{figure*}

The distribution of the included participants among the categories levels of {\em Education\/}, {\em Statistician\/}, {\em Age\/}, and {\em Gender\/} are displayed in Figure~\ref{PPhr_fig:HistorgramsDistributionCatVar}. 
These bar plots show that the number of statisticians is much lower than the number of non-statisticians (226 vs.\ 655), as can be expected. On the other hand, the participants are evenly distributed among the genders (430 male vs.\ 440 female) and the middle two age groups (363 of 20-39 years and 375 of 40-60 years). There were more males than females among the statisticians (133 vs.\ 89 respectively).

Note that many of the participants were highly educated. This is partially explained by the fact that most statisticians are highly educated. However, even among the non-statisticians, the number of highly educated persons is high (58\%). This disproportionally large number of highly educated participants is simply due to the fact that the non-statisticians are a convenience sample. 

\newpage

\section{Results}

The distributions of the interpreted percentages of each probability phrase are displayed by the density plots in Figure~\ref{PPhr_fig:DensityPlotsProbabilityPhrases} and the mean values are listed on the right side of the plots. For most phrases the numerical interpretations have a range with a width of 30 to 40 percentage points. For example, the numerical interpretations of {\em sometimes\/}, {\em probable\/}, and {\em almost always\/} respectively range from 5-55\%, 40-95\%, and 70-100\%. These large widths clearly indicate that the interpretations of probability phrases differ a lot among people. 

Even though there is a lot of variability, there seems to be some consensus about the interpretation of extreme words like {\em always\/}, {\em certain\/}, {\em never\/}, and {\em impossible\/}. Namely, the widths of these plots is only 15 percentage points, with the most frequently chosen percentages for {\em always\/} and {\em certain\/} between 95\% and 100\%, and between 0\% and 5\% for {\em never\/} and {\em impossible\/}. The higher agreement for these words is in accordance with previous research and to be expected, since these are phrases with strong meanings which leave little room for interpretation.

Another thing to notice is that, given the peaks in the density plots, many people seem to prefer to express probabilities as multiples of ten. Also, there seems to be no phrase in this list that represents 50\%. The candidates {\em sometimes\/} to {\em possible\/}, for which 50\% is the most frequently chosen interpretation, all have a large tail on the left. Hence, if used to indicate 50\%, they will probably be underestimated by many people and only a few will overestimate the intended percentage.

\afterpage{
\begin{figure*}[h!]
	\centering
	\includegraphics[width=\textwidth]{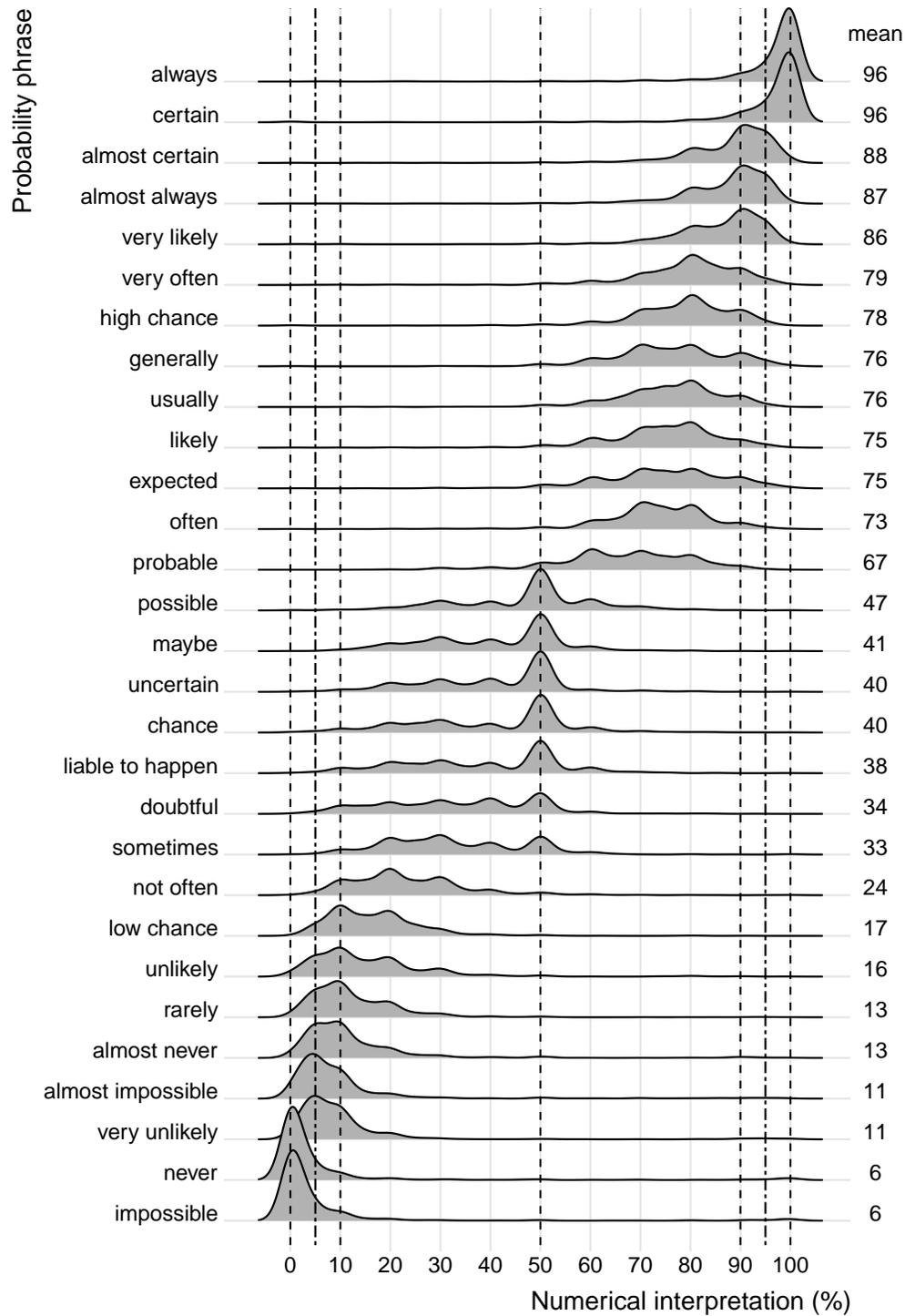}
	\caption{Density plots and mean values of the numerical interpretations (in percentages) given by all participants for each phrase in the survey. Note that density plots are a smooth variant of histograms and may therefore be positive outside the data range of 0-100\%.}
	\label{PPhr_fig:DensityPlotsProbabilityPhrases}
\end{figure*}
\clearpage
} 

\subsubsection*{Comparison with previous research}
The ten Dutch frequency phrases included in this study were also in the study by \citet{Art:DoorgaansDikwijls_EekhofMolPielage}. For nine of these phrases, the mean values associated with it in these two studies differed a maximum of three percentage points. Only the interpretation of {\em sometimes\/} differed more, namely a difference of 8 percentage points (mean of 33\% in our study vs.\ 25\% in their study). 

There are many English studies that included English counterparts of the Dutch phrases in our study. \citet{Art:Meta-Analysis_Theil} listed the mean interpretations for ten probability phrases found in ten studies. Seven of these phrases are also included in our study. The mean interpretations that we measured are all between the lower and upper bounds of the means measured in these ten studies. However, these ranges were quite wide for some expressions, indicating that there is too much variability in the English studies to conclude from this result that there are no differences between the interpretations of Dutch phrases and their English translations.

\subsubsection*{Synonyms and asymmetry}

As mentioned in the introduction, some expressions can be considered to be synonyms. For example, \citet{Art:QuantMeaningProbExpr_ReaganYoutz1989} concluded from their distribution plots that, amongst others, {\em likely\/} and {\em probable\/}, and {\em low chance\/} and {\em unlikely\/} are synonymous. The density plots from our data visually confirm that the Dutch translations of these expressions are synonymous as well. Other examples of synonymous pairs are {\em almost always\/} and {\em very likely\/}, and {\em doubtful\/} and {\em sometimes\/}, and more pairs are suggested by 	Figure~\ref{PPhr_fig:DensityPlotsProbabilityPhrases}.

Given that {\em low chance\/} and {\em unlikely\/} are synonymous, one might expect {\em high chance\/} and {\em likely\/} to be synonymous as well. However, the density plot of {\em high chance\/} in Figure~\ref{PPhr_fig:DensityPlotsProbabilityPhrases} shows that its interpretation is actually in between those of {\em likely\/} and {\em very likely\/} (i.e.\ {\em likely\/}'s interpretation is usually somewhat lower and {\em very likely\/}'s interpretation peaks at a higher percentage). This suggests that there is asymmetry in the interpretations of Dutch probability phrases. However, given that \citet{Art:QuantMeaningProbExpr_ReaganYoutz1989} concluded that {\em high chance\/} and {\em likely\/} are synonyms, the asymmetry may be different in Dutch and English.  

The imbalance in the interpretation of mirrored probability phrases is often investigated by reviewing whether the group means or group medians of the interpretations of two complementary words sum to 100\%. For instance, \citet{Art:EmpiricalScaling_LichtensteinNewman} concluded that the interpretations of {\em likely\/} and {\em unlikely\/} are asymmetric, since their means sum to (72\% + 18\% =) 90\% and their medians sum to (75\% + 16\% =) 91\%. This assymmertry was confirmed by both \citet{Art:QuantMeaningProbExpr_ReaganYoutz1989} (medians sum to 90\%) and \citet{Art:BlindChance_Stheeman} (medians sum to 80\%). Our data shows that the assymmetry is also present for the Dutch translations of {\em likely\/} and {\em unlikely\/}. Namely, the mean interpretation of {\em likely\/} in our data is 75\% (see means listed in Figure~\ref{PPhr_fig:DensityPlotsProbabilityPhrases}) and the mean for {\em unlikely\/} is 16\%, and hence these sum to 91\%. 

Previous studies have also shown that some terms actually are (almost) symmetrical. For example, {\em very likely\/} and {\em very unlikely\/} (mean interpretations sum to 96\% \citep{Art:EmpiricalScaling_LichtensteinNewman}), and {\em almost always\/} and {\em almost never\/} (median interpretations sum to 98\% \citep{Art:BlindChance_Stheeman}). We found these symmetries also for the Dutch counterparts of these complementary phrases (means sum to, respectively, 97\% and 100\%). 

This method used to find (a)symmetry in the data actually only studies this phenomenon on the group level; the means or medians of all data are compared. To investigate asymmetry on an individual level, we propose to sum the interpretations of complementary phrases for each individual and look at the distribution of this sum. If the distribution of these sums is positive at 100\%, this indicates that the expressions are interpreted symmetrically by some people. Results that sum to less or more than 100\% indicate asymmetrical interpretations.

\begin{figure*}[h!]
	\centering
	\includegraphics[width=0.7\textwidth]{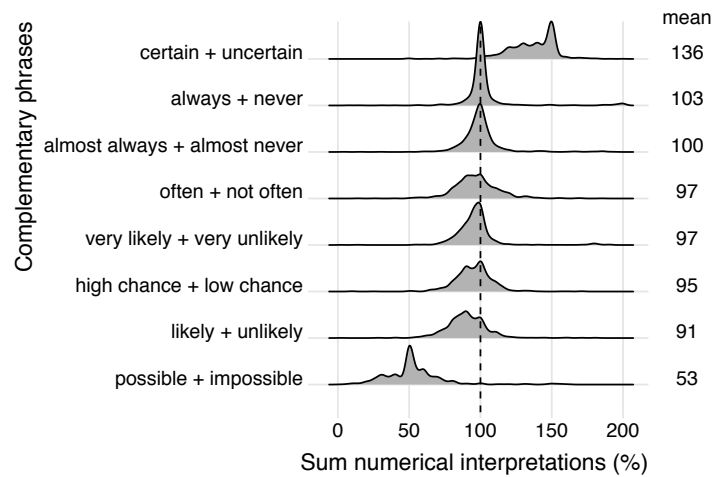}
	\caption{Density plots and mean values of the sums of the numerical interpretations (in percentages) given by all participants for the complementary phrase pairs in the survey.}
	\label{PPhr_fig:DensityPlotsComplentaryPhrases}
\end{figure*}

 Density plots of the sums of complementary phrases investigated in this study are shown in Figure~\ref{PPhr_fig:DensityPlotsComplentaryPhrases}. These plots show that there are some mirrored pairs which interpretation sums up to about 100\% for most participants, for example {\em (almost) always\/} and {\em (almost) never\/}, and {\em very likely\/} and {\em very unlikely\/}. Other complementary phrases were interpreted asymmetrically by many participants and usually sum up to slightly less than 75\% to 100\%, e.g.\ {\em likely\/} and {\em unlikely\/}, and {\em often\/} and {\em not often\/}.

Our data confirms that the conclusion of \citet{Art:QuantifyingProbExpr_MostellerYoutz}, regarding the very strong asymmetry in the interpretation of {\em possible\/} and {\em impossible\/}, can be generalized to the Dutch translations of these expressions. Namely, Figure~\ref{PPhr_fig:DensityPlotsProbabilityPhrases} shows that {\em impossible\/} indeed has a stable interpretation that is close to 0\%, while {\em possible\/} has a broad interpretation from 20\% to 70\% which peaks around 50\%. The asymmetry is also confirmed by the distribution of their sums in Figure~\ref{PPhr_fig:DensityPlotsComplentaryPhrases}. 

Another mirrored pair that shows a strong asymmetry in the interpretation is {\em certain\/} and {\em uncertain\/}. There is a consensus on the interpretation of {\em certain\/} (around 100\%) while the perception of {\em uncertain\/} varies a lot and is comparable to {\em maybe\/}'s interpretation, i.e.\ some value between 20\% to 50\% (see Figure~\ref{PPhr_fig:DensityPlotsProbabilityPhrases}). As a result, the percentages of {\em certain\/} and {\em uncertain\/} always sum to more than 100\% and together peak at 150\% (Figure~\ref{PPhr_fig:DensityPlotsComplentaryPhrases}).

\subsubsection*{Context}
One of our concerns was that the context of the sentences influences the perception of the probability phrases. To avoid the base-rate effect, we tried to formulate the context sentences as neutral as possible. 

\begin{figure*}[h!]
	\centering
	\includegraphics[width=\textwidth]{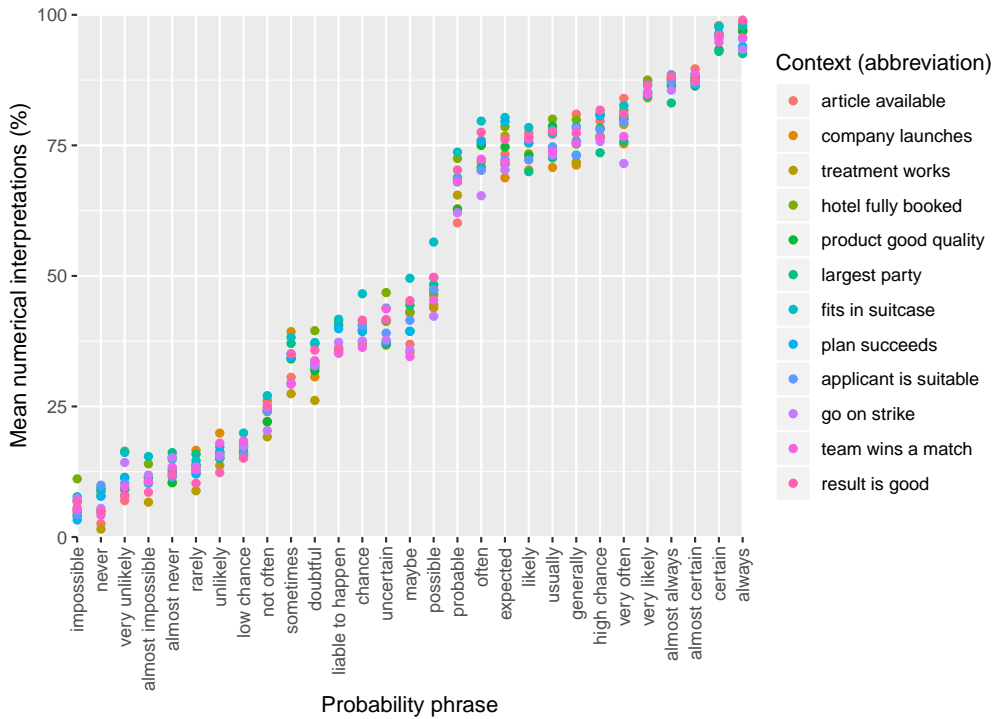}
	\caption{Means of the numerical interpretations (in percentages) given by all participants for each phrase in the surveys, grouped by the context of the sentences. Listed contexts in the legend are abbreviations of the originals (see Table~\ref{PPhr_tab:ContextSentencesUsed} in Appendix~\ref{PPhr_app:ContextPhrases}).}
	\label{PPhr_fig:MeanPercentagesPerPhraseByContext}
\end{figure*}

To check whether we succeeded in our intention, we investigated the variability of the interpretation of phrases among different contexts. Figure~\ref{PPhr_fig:MeanPercentagesPerPhraseByContext} shows the mean percentages given by the participants to each probability phrase, grouped by context. This plot shows that, in general, the means of a phrases are very similar for each context, with a maximum of 25 percentage points difference between contexts. Most of this variability appears for words that represent 30\% to 80\%. 

Most importantly, although the plots show some influence of context on the interpretations, they do not suggest that any of the sentences is systematically interpreted differently (higher/lower or more/less extreme) from the others. This lack of systematic difference suggests that the chosen contexts were neutral enough to prevent a strong base-rate effect. 

\subsubsection*{Differences between sub-populations}

\begin{figure*}[h!]
	\centering
	\includegraphics[width=\textwidth]{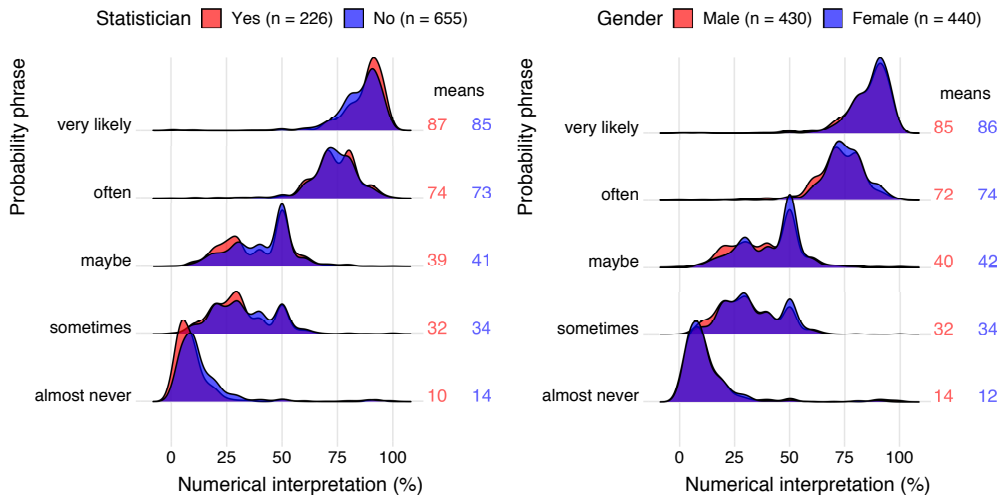}
	\caption{Density plots and mean values of the numerical interpretations (in percentages) given by statisticians and non-statisticians, and males and females for a selection of five phrases from the survey. Gender category level {\/\em Other/Prefer not to say} was omitted, because there were only 11 observations in this group. Note that density plots are a smooth variant of histograms and may therefore be positive outside the data range of 0-100\%.}
\label{PPhr_fig:ComparisonGroups_Gender_Statistician}
\end{figure*}

One of the aims of this research was to make a comparison of the interpretation of probability phrases of different sub-populations, namely to compare interpretations of experts (statisticians) with those of laymen, and also males with females. Figure~\ref{PPhr_fig:ComparisonGroups_Gender_Statistician} shows the density plots of the statisticians and non-statistician (left panel) and those for men and women (right panel) for a selection of five probability phrases. These expressions were selected from different ranges of numerical interpretations. The results for all phrases are shown in Figure~\ref{PPhr_fig:DensityPlots_StatisticiansVSNonStatisticians_AllData} and Figure~\ref{PPhr_fig:DensityPlots_MaleVSFemale_AllData} in Appendix~\ref{PPhr_app:ComparingGroups_AllPhrases}.

These density plots show that the interpretations of the probability phrases are very similar for both statisticians and non-statisticians, and for both genders. This similarity is represented by the overlapping regions of the plots. The non-overlapping regions are relatively small, which suggests that there are no big differences between the groups. This is supported by the group means, since the maximum difference between the genders and between statisticians and non-statisticians is four percentage points.

Although the differences are small, the density plots of {\em verly likely\/} and {\em almost never\/} in the left panel of Figure~\ref{PPhr_fig:ComparisonGroups_Gender_Statistician} may suggest that statisticians are more extreme in their judgements of these words than non-statisticians. This phenomenon is also seen for other extreme phrases (see Figure~\ref{PPhr_fig:DensityPlots_StatisticiansVSNonStatisticians_AllData} in Appendix~\ref{PPhr_app:ComparingGroups_AllPhrases}), but not for phrases expressing percentages closer to 50\%. However, the difference between the group means is small for these phrases, so the group effect (if present) is weak.

\section{Conclusions and discussion}

In this study we have investigated the variability of the interpretation of Dutch probability and frequency phrases. The set-up of our survey was comparable to previous surveys on the interpretation of English phrases, but it filled some gaps in the research on Dutch probability phrases. For example, we included many Dutch expressions that were not studied before and represented them in a neutral context. Furthermore, we investigated synonymous words, asymmetries in the interpretation of mirrored phrases, and differences in interpretation between statisticians and non-statisticians, and between males and females. 

Our results showed that, as in English, there is a large variability in the interpretation of Dutch probability and frequency phrases. Although there is some agreement about extreme words as {\em always\/}, {\em certain\/}, {\em never\/}, and {\em impossible\/}, there is no consensus about words that describe a less extreme probability. The mean interpretations of the Dutch phrases are within the range of means found for their English translations in other studies. However, those results varied a lot, so stating that the meaning of a phrase is maintained after translating it from Dutch to English (or visa versa) would be a too strong conclusion.

The data indicated that some phrases may be synonymous, since they showed a comparable range and mean of interpretations. However, the interpretations of these synonyms is inconsistent. We also found asymmetry in the interpretation of phrases. For example, usually an individual's numerical interpretations of {\em likely\/} and {\em unlikely\/} do not sum to 100\%. This asymmetry is larger if one of the complementary phrases can have different meanings, like expressions as {\em possible\/} and {\em uncertain\/}. English studies showed comparable asymmetrical patterns. 

To test for differences in interpretations between experts and laymen this survey was distributed among both statisticians and non-statisticians. Our data showed large variability within each group and no structural differences between them. Hence, it seems that regularly making and communicating probability estimations does not increase agreement about the interpretations of probability expressions. We also found no differences between the interpretations of male and female participants.

One of the strengths of this study is its large sample size of 881 participants. In most studies sample sizes were quite small, for instance the the number of participants in the Dutch and English studies listed in this article ranged from 7 to 238, with mean 86.8. \citet{Art:Meta-Analysis_Theil} mentioned one study with a larger sample size, namely the research by \citet{Art:HeterogeneousSample_ClarkeEtAl} which included 966 respondents. 

A limitation of our research is that our sample is largely a convenience sample, since non-statisticians were invited to participate via Twitter \citep{Twitter}. This probably let to the disproportionate distribution of education levels and, hence, our results on the non-statisticians may not generalize to the population. However, the results from this study are still valuable, since they showed that, even within this homogeneous sample, interpretations of probability expressions differed enormously. This indicates that interpretation are dissimilar even among like-minded person. If the sample had been more heterogeneous, the interpretations would probably have varied even more. 

For a follow-up study, researchers may consider to allow participants to give a lower and upper bound of their numerical interpretation of a phrase. In our survey, participants were asked to indicate their interpretation of an expression as a point estimate on the interval from 0\% to 100\%. This approach is a clear and simple and satisfactory for most phrases. However, these single numerical values do not give any information about the interpretation range on an individual level. Furthermore, ranges may be helpful to investigate the different types of interpretations of words like {\em possible\/} and {\em uncertain\/}.

Based on this study, we conclude that making a translation table from Dutch verbal probability phrases to numeric values is infeasible. Many expressions, both with and without adverbs, have too variable interpretations to make such a codification. The strong variability within sub-populations indicates that even group-specific codifications are not feasible. 

The asymmetry in the interpretation of mirrored terms also indicates one of the complications when making a translations table. Namely, if a table was made to translate verbal probability phrases into numerical values, the codification would be asymmetrical. Asymmetry in such a table would be counter-intuitive and impractical. 

Additionally, although the context sentences used in our research did not show strong base-rate effects, it did indicate that context has some influence on the interpretations. This effect of context shows another complication for making a translation table between verbal and numeric probability expressions; a different table would required for each situation. 

Hence, we conclude that reporting probabilities verbally may lead to serious miscommunication, even between people from the same sub-population. Therefore, our advice is to be more precise if you want to convey a probability estimate and to consider using a numerical instead of a verbal expression, or at least both. This will also prevent the intended probability from getting lost in its translation from one language to another. If desired, you can also include uncertainty in numerical statements by using phrases like {\em there is at least 80\% chance of rain\/} or {\em roughly one third of the voters support this policy\/}.


\bibliographystyle{apa}
\bibliography{./Bibliography_ProbPhrases.bib}

\newpage
\appendix
\section{Translations of probability phrases}\label{PPhr_app:TranslationProbabilityPhrases}

\begin{table}[h!]
		\caption{The 29 Dutch frequency and probability phrases used in the survey, with their English translations used in this article. The phrases are presented in the same order as in Figure \ref{PPhr_fig:DensityPlotsProbabilityPhrases} in this article.}
	\centering
	\begin{tabular}{ll}
		\hline
		Dutch phrase & English translation \\ 
		\hline
		onmogelijk & impossible \\ 
		nooit & never \\ 
		zeer onwaarschijnlijk & very unlikely \\ 
		bijna onmogelijk & almost impossible \\ 
		bijna nooit & almost never \\ 
		zelden & rarely \\ 
		onwaarschijnlijk & unlikely \\ 
		kleine kans & low chance \\ 
		niet vaak & not often \\ 
		soms & sometimes \\ 
		twijfelachtig & doubtful \\ 
		kan gebeuren & liable to happen \\ 
		kans & chance \\ 
		onzeker & uncertain \\ 
		misschien & maybe \\ 
		mogelijk & possible \\ 
		vermoedelijk & probable \\ 
		vaak & often \\ 
		te verwachten & expected \\ 
		waarschijnlijk & likely \\ 
		meestal & usually \\ 
		doorgaans & generally \\ 
		grote kans & high chance \\ 
		heel vaak & very often \\ 
		zeer waarschijnlijk & very likely \\ 
		bijna altijd & almost always \\ 
		bijna zeker & almost certain \\ 
		zeker & certain \\ 
		altijd & always \\ 
		\hline
	\end{tabular}
	\label{PPhr_tab:TranslationProbabilityPhrases}
\end{table}

\newpage
\section{Context sentences}\label{PPhr_app:ContextPhrases}

\begin{table}[h!]
	\caption{The 12 Dutch context sentences used in the survey, with their English translations used in this article.}
	\label{PPhr_tab:ContextSentencesUsed}
	\begin{tabular}{l}
		\hline
		Dutch context sentences and their English translations (italic) \\
		\hline
		Het is waarschijnlijk dat alles in de koffer past. \\ \quad \em It is likely that everything fits in the suitcase. \\
		Het is waarschijnlijk dat het team een wedstrijd wint. \\ \quad \em  It is likely that the team wins a match. \\
		Het is waarschijnlijk dat deze behandeling aanslaat.  \\ \quad \em  It is likely that this treatment will work. \\
		Het is waarschijnlijk dat een sollicitant geschikt is voor de baan. \\ \quad \em  It is likely that this applicant is suitable for the job.  \\
		Het is waarschijnlijk dat bedrijf A het product eerder lanceert dan bedrijf B. \\ \quad \em  It is likely that company A launches the product before company B does. \\
		Het is waarschijnlijk dat zij gaan staken.  \\ \quad \em  It is likely that they will go on strike.\\
		Het is waarschijnlijk dat de uitslag goed is. \\ \quad \em  It is likely that the result is good. \\
		Het is waarschijnlijk dat deze partij de grootste wordt bij de verkiezingen. \\ \quad \em  It is likely that this party will be the largest in the elections. \\ 
		Het is waarschijnlijk dat dit plan slaagt. \\ \quad \em  It is likely that this plan succeeds. \\
		Het is waarschijnlijk dat deze producten van goede kwaliteit zijn. \\ \quad \em  It is likely that these products are of good quality. \\
		Het is waarschijnlijk dat dit hotel is volgeboekt. \\ \quad \em  It is likely that this hotel is fully booked.\\
		Het is waarschijnlijk dat dit artikel verkrijgbaar is.  \\ \quad \em  It is likely that this article is available.\\
		\hline
	\end{tabular}
\end{table}

\newpage
\section{Differences between sub-populations}\label{PPhr_app:ComparingGroups_AllPhrases}

\begin{figure*}[h!]
	\centering
	\includegraphics[width=\textwidth]{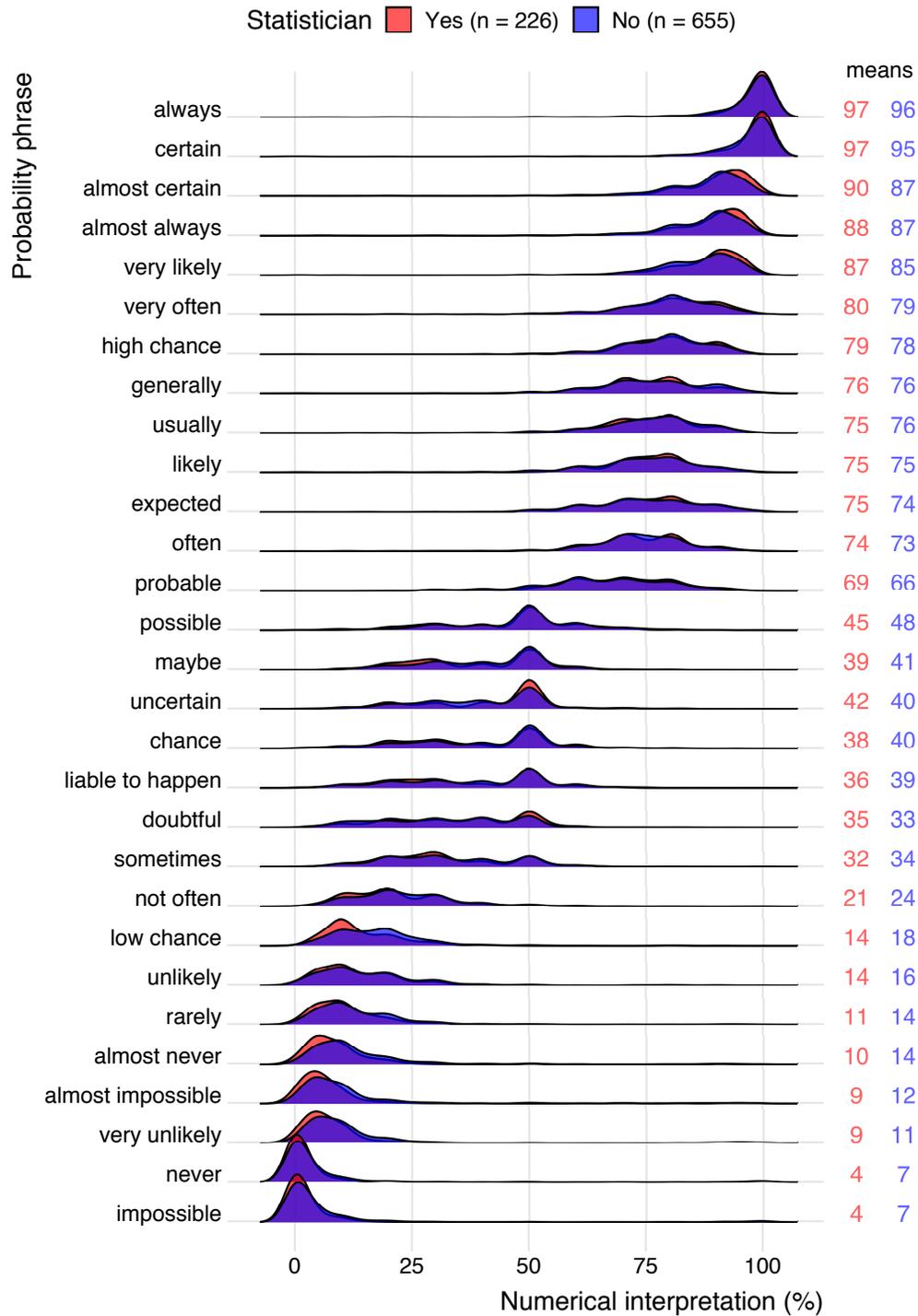}
	\caption{Density plots and mean values of the numerical interpretations (in percentages) given by statisticians and non-statisticians for each phrase in the survey. Note that density plots are a smooth variant of histograms and may therefore be positive outside the data range of 0-100\%.}
	\label{PPhr_fig:DensityPlots_StatisticiansVSNonStatisticians_AllData}
\end{figure*} 

\begin{figure*}[h!]
	\centering
	\includegraphics[width=\textwidth]{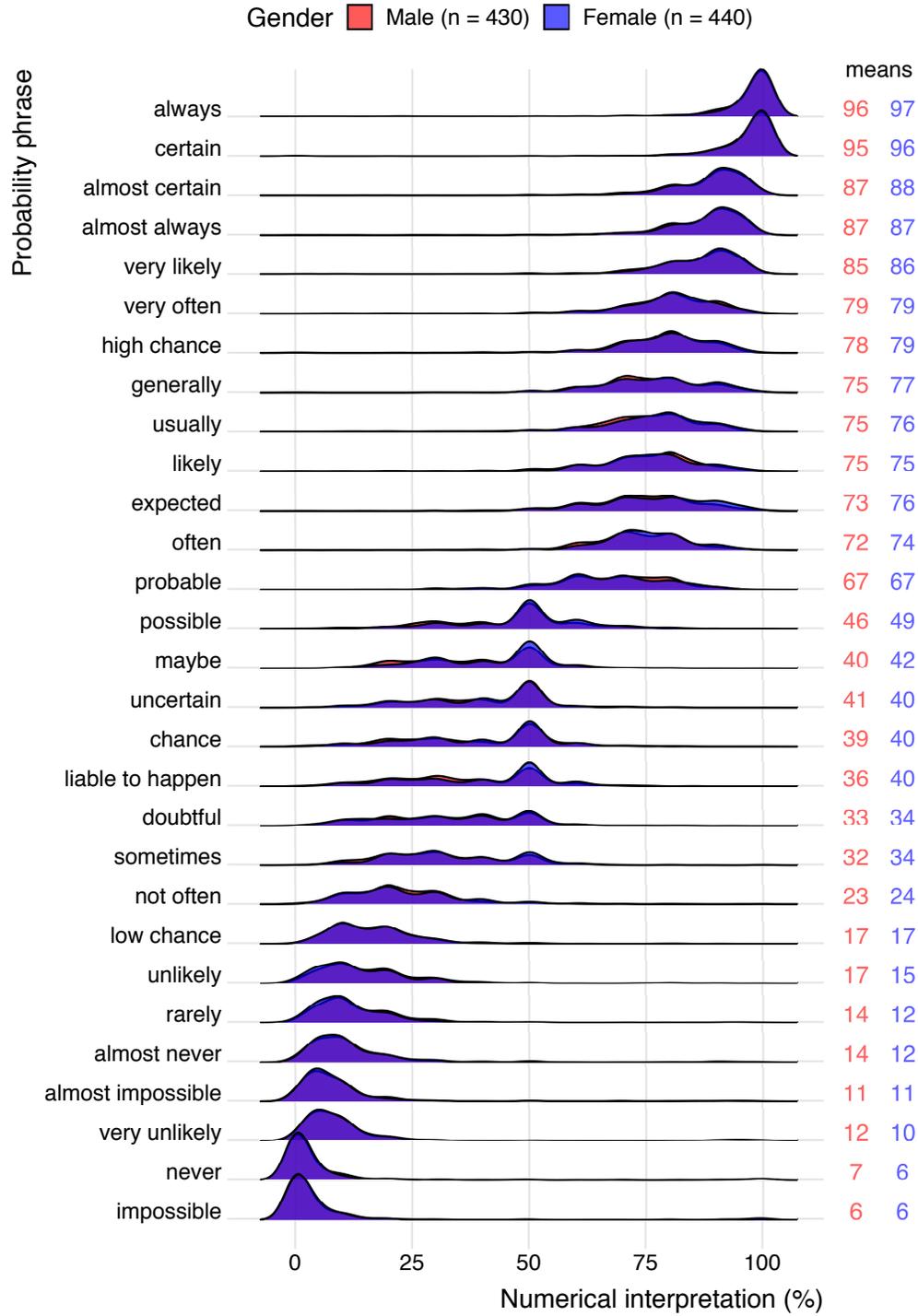}
	\caption{Density plots and mean values of the numerical interpretations (in percentages) given by males and females for each phrase in the survey. Category level {\/\em Other/Prefer not to say} was omitted, because there were only 11 observations in this group. Note that density plots are a smooth variant of histograms and may therefore be positive outside the data range of 0-100\%.}
	\label{PPhr_fig:DensityPlots_MaleVSFemale_AllData}
\end{figure*}

\newpage
\section{Supplementary material}

The data and a .rmd-file with all analyses are provided online as supplementary material.

\end{document}